\documentclass[11pt,a4paper]{amsart}

\usepackage{amsmath,amsfonts,amssymb}
\usepackage{mathrsfs}

\usepackage{cases}

\numberwithin{equation}{section}

\usepackage{amsthm}
\theoremstyle{plain}
\theoremstyle{definition}
\newtheorem*{example}{Example}
\theoremstyle{remark}
\newtheorem*{remark}{Remark}

\newcommand{\D}{\mathop{}\!\mathrm{d}}

\newcommand{\I}{{\mathrm{i}\mkern1mu}}

\newcommand{\E}{{\mathrm{e}}}

\newcommand{\Pfrac}[3][]{\frac{\partial^{#1} #2}{\partial {#3}^{#1}}}

\DeclareMathOperator{\sgn}{sgn}

\newcommand{\bra}[1]{\langle #1 |}
\newcommand{\ket}[1]{| #1 \rangle}

\newcommand{\Gph}{G_{\text{caus}}}
\newcommand{\G}{G_{\text{st}}}

\newcommand{\Th}{K}

\newcommand{\tc}{\tilde{c}}

\newcommand{\dc}{\partial_c}

\newcommand{\Qi}{Q_{\text{ind}}}
\newcommand{\Vi}{V_{\text{ind}}}

\newcommand{\SL}{S_{\text{L}}}
\newcommand{\SR}{S_{\text{R}}}
\newcommand{\SLR}{S_{\text{LR}}}

\newcommand{\y}{\psi}
\newcommand{\dy}{\partial_\psi}

\newcommand{\dx}{\partial_x}

\begin{document}

\title{New Objects in Scattering Theory with Symmetries}

\author{Andrey S. Losev}

\address{Wu Wen-Tsun Key Lab of Mathematics, Chinese Academy of Sciences}
\address{National Research University Higher School of Economics, Moscow, Russia
}
\email{aslosev2@yandex.ru}

\author{Tim V. Sulimov}

\address{Petersburg Department of the Mathematical Institute of the Russian
Academy of Science, St. Petersburg}
\email{optimus260@gmail.com}

\maketitle

\date{\today}

\begin{abstract}
	We consider 1D quantum scattering problem for a Hamiltonian with symmetries.
	We show that the proper treatment of symmetries in the spirit of homological
	algebra leads to new objects, generalizing the well known T- and K-matrices.
	Homological treatment implies that old objects and new ones are be
	combined in a differential. This differential arises from homotopy transfer
	of induced interaction and symmetries on solutions of free
	equations of motion.  Therefore, old and new objects satisfy remarkable
	quadratic equations. We construct an explicit example in SUSY QM on $S^1$
	demonstrating nontriviality of the above relation.
\end{abstract}


\section{Introduction and Results}%
\label{sec:introduction_and_results}

Consider nonrelativistic 1D quantum mechanics for a particle that has internal
degrees of freedom, e.g. spin or isospin. Its space of states is described by a
tensor product of  functions on a circle or a line and a finite dimensional
vector space $I$ corresponding to internal degrees of freedom. The Hamiltonian
is taken to be $H = H_0 + V$, where $H_0$ is the \emph{free Hamiltonian} that
acts by $-\dx^2$ and does not rotate the internal space $I$, and the
\emph{interaction} $V$ that in general rotates the space $I$. We shall use the
perturbative approach, i.e., start with a solution to $H_0$ and consider the
corresponding solution of $H$ as a series in interaction $V$. Since we will be
interested in solutions of $H_0$ with fixed energy $E$, it is convenient to
define a shifted Hamiltonian $H_{0,E} = H_0 - E$.

Consider the problem on a line and assume that interaction falls off fast enough
at space infinities. Define the \emph{perturbative T-matrix} as follows. Its
matrix element between two states of the continuous spectrum
$\ket{\varphi_\alpha}, \ket{\varphi_\beta} \in \ker{H_{0,E}}$ is equal to
\begin{align}\label{Tba}
	T_{\beta,\alpha} = \bra{\varphi_\beta} T \ket{\varphi_\alpha}
	= \bra{\varphi_\beta} V + V \Gph V + V \Gph V \Gph V + \ldots
	\ket{\varphi_\alpha},
\end{align}
where $\Gph$ is the \emph{causal Green's function}
\begin{equation}\label{G_ph}
	\Gph = \lim_{\varepsilon \to 0+} (E - H_0 + \I \varepsilon)^{-1}.
\end{equation}
It is known that $T$ is the perturbative solution to the Lippman-Schwinger
equation for the (nonperturbative) T-matrix $T^{\text{np}}$:
\begin{equation}\label{LS_T}
	T^{\text{np}} = V + V \Gph T^{\text{np}}.
\end{equation}
It is also well-known that all questions about scattering of particles may be
expressed through $T^{\text{np}}$ (see, for example,~\cite{taylor1972}).

Let $\mathfrak{g}$ be a symmetry (super)algebra of $H$ with generators $S_a$:
\begin{equation}
	[S_a, H] = 0, \qquad \{S_a, S_b\}= f_{ab}^{d} S_{d},
\end{equation}
where $\{ ,\}$ denotes the supercommutator.
Note, that Hamiltonian itself is also a symmetry, because $[H, H] = 0$.
One is prompted to ask if the symmetries of the Hamiltonian are reflected in
any kind of symmetries of $T^{\text{np}}$, or at least in symmetries of $T$.

In this paper we will attempt to answer this question. Surprisingly, the answer
is very nontrivial. In order to formulate it we have to introduce \emph{new
objects} that (to the best of our knowledge) have not been introduced in
scattering theory before. We will also show that they are related by quadratic
equations. In our current understanding these objects may be seen as coming from
the problem on a circle of asymptotically big radius. However, we believe that
they could be defined directly in the theory on the line, we just do not know of
such a definition yet and we are working on it.

To formulate the answer we will use \emph{ghosts} for symmetries. The ghosts
$c^a$ are coordinates on $\mathfrak{g}$ with inverted parity. In simple words it
means that if $S_a$ is even then $c^a$ is odd and vice versa. The ghost for the
Hamiltonian will be denoted by $c$ and will be odd while the rest of the ghosts
will be denoted by $\tc^a$.

It turns out that $T$ is just a particular component of the differential
$\Qi$ acting on the space $\ker{H_{0,E}} \otimes \mathbb{C}[[c^a, R^{-1}]]$. 
Here  $R^{-1}$ is an even variable. We call it $R^{-1}$ because it is related to
the problem on the circle of radius $R$ which is discussed in
subsections~\ref{sub:double_delta_potential_on_s_1_}
and~\ref{sub:_r_to_infty_limit}.

\emph{Appearance of the differential $\Qi$ is novel in the context of scattering
theory, but is natural in the context of homological algebra.}

Quadratic equations are just 
\begin{equation}\label{Qi2}
	\Qi^2=0.
\end{equation}

Now we will proceed to details of construction of $Q$ out of symmetries.
First we assume that like Hamiltonian symmetries $S_a$ are split into two pieces:
\begin{equation}
	S_a=S_{0,a}+S_{1,a}
\end{equation}
where $S_{0,a}$ is a symmetry of the free Hamiltonian and $S_{1,a}$ (like
interaction of Hamiltonian) vanish at space infinities.

Then,  
\begin{equation}\label{Qi_in_c}
	\Qi = \tfrac{1}{2} f_{ab}^{d} c^a c^b \partial_{c^d} + c^a S_a (R^{-1}) +
	c^a c^b S_{ab}(R^{-1}) + \ldots
\end{equation}

In the zeroth order in $R^{-1}$ expansion the second term in equation above
equals to the symmetry of free Hamiltonian acting on solutions to free problem
\begin{equation}
	\Qi^{(0)} = \tfrac{1}{2} f_{ab}^{d} c^a c^b \partial_{c^d} + c^a
	S_{0,a}\big|_{\ker{H_{0,E}}}.
\end{equation}
In the first order in $R^{-1}$ expansion we have
\begin{equation}
	\Qi^{(1)} =  c^a S_a^{(1)} +c^a c^b S_{ab}^{(1)}+\ldots = c K + \tc^a
	S_a^{(1)} +c^a c^b S_{ab}^{(1)}+\ldots
\end{equation}

It is remarkable that the coefficient in front of the distinguished ghost $c$ is
the well known K-matrix (see, for instance,~\cite{newton1982}), which is related
to T-matrix by 
\begin{equation}\label{Th_T}
	\Th = T + \frac{\I}{2 \sqrt{E}} T \Th.
\end{equation}

\begin{example}
	Consider $V(x) = \lambda \delta(x)$. The T-matrix can be easily acquired
	through any of the formulae above giving
	\begin{equation}
		T(p, q) = \frac{\lambda}{1 - \frac{\lambda}{2 \I \kappa}},
	\end{equation}
	which is independent of the incoming or outgoing momenta. Using~\eqref{Th_T}
	we get
	\begin{equation}
		\Th(p,q) = \lambda.
	\end{equation}
	Coincidently, $\Th$ turned out to be equal to $V$. One can easily see that
	this will not be the case in general.
\end{example}

The perturbative expansion for $K$ can be acquired from~\eqref{Tba} by replacing
the causal Green's function with the standing wave one.

\emph{Other terms in the expansion of $\Qi$ are novel objects in scattering
theory, related to standard objects by $\Qi^2 = 0$.}

\section{Theory}%
\label{sec:theory}

\subsection{Homotopy Transfer}%
\label{sub:homotopy_transfer}

In this paper we shall illustrate ideas of
section~\ref{sec:introduction_and_results} using the simplest nontrivial example
of $\mathscr{N} = 1$ supersymmetric quantum mechanics on $S^1$ of radius $R$.
Moreover, for simplicity we consider an algebra consisting of the Hamiltonian
together with a single odd symmetry that squares to zero:
\begin{equation}\label{sym_H}
	[S, H] = 0, \qquad S^2 = 0.
\end{equation}
Thus, all the structure constants $f_{ab}^d$ are zero, so the first term
in~\eqref{Qi_in_c} is also zero.

The corresponding ghosts are $c$ and $\tc$ with the following
properties:
\begin{enumerate}
	\item $c$ is odd while $\tc$ is even;
	\item $c$ anti-commutes with $S$.
\end{enumerate}
These ghosts allow us to construct differentials
\begin{equation}\label{diff}
	Q_0 = c H_{0,E} + \tc S_0, \quad Q_1 = c V + \tc S_1, \quad Q = Q_0 + Q_1,
\end{equation}
and a \emph{homotopy}
\begin{equation}\label{hom}
	h = \G \dc, \quad \G \ket{\varphi} =
	\begin{cases}
		(E - H_0)^{-1} \ket{\varphi}, &\ket{\varphi} \notin \ker{H_{0,E}},\\
		0, &\ket{\varphi} \in \ker{H_{0,E}},
	\end{cases}
\end{equation}
where $\dc = \Pfrac{}{c}$.
In physics $\G$ is known as the standing wave Green's function, but in the
spirit of homological algebra we will refer to it as the \emph{homotopical}
Green's function.

From~\eqref{sym_H} and the properties of the ghosts it follows that
\begin{equation}\label{Q02Q2}
	Q_0^2 = Q^2 = 0.
\end{equation}

Let $\mathscr{H}$ denote the original Hilbert space. Define $U = \mathscr{H}
\otimes \mathbb{C}[c,\tc]$, then the notation is as follows
\begin{equation}\label{pi_i}
	U = \ker{H_{0,E}} \oplus U_{\text{ac}}, \quad \pi : U \to \ker{H_{0,E}},
	\quad i : \ker{H_{0,E}} \to U,
\end{equation}
where $\pi$ and $i$ are the projection and inclusion operators. Composing them
we get the projection on $\ker{H_{0,E}}$ that can be written using the homotopy
\begin{equation}\label{i_pi}
	i \pi = \Pi_{\ker{H_{0,E}}} = 1 + \{ h, Q_0 \}.
\end{equation}

We can now write down the central object of homotopy transfer~--- the
\emph{induced differential} (see, for example,~\cite{losev2019}
and~\cite{arvanitakis2020})
\begin{equation}
	\Qi = \pi Q_0 i + \pi Q_1 i + \pi Q_1 h Q_1 i + \pi Q_1 h Q_1 h Q_1 i +
	\ldots.
\end{equation}
A brief calculation gives
\begin{equation}
	\Qi = \tc S_0 + c \Vi + c \dc \tc \SR + \dc c \tc \SL - \dc \tc^2 \SLR,
\end{equation}
where $S_0$ is actually $S_0\big|_{\ker{H_{0,E}}}$, but since $[S_0, H_0] = 0$
we simplify the notation. Additionally,
\begin{equation}\label{V_ind}
	\Vi = \pi \left( V + V \G V + V \G V \G V + \ldots \right) i,
\end{equation}
\begin{equation}\label{S_R}
	\SR = \pi \left( S + V \G S + V \G V \G S + \ldots \right) i,
\end{equation}
\begin{equation}\label{S_L}
	\SL = \pi \left( S + S \G V + S \G V \G V + \ldots \right) i,
\end{equation}
\begin{equation}\label{S_LR}
	\SLR = \pi \left( S \G S + S \G V \G S + S \G V \G V \G S + \ldots \right)i.
\end{equation}

We can now use the main result of homotopy transfer. Since $Q^2 = 0$, by
Kadeishvili's theorem we have
\begin{equation}\label{mc_ind}
	\Qi^2 = 0.
\end{equation}
This equation can be expanded as
\begin{equation}
\begin{split}
	\Qi^2 &= c \tc [\Vi, S_0] + c \dc \tc^2 \{ S_0, \SR \}
	+ \dc c \tc^2 \{ S_0, \SL \} + \dc \tc^3 [S_0, \SLR] \\
	&\quad+c \tc \left( \Vi \SL - \SR \Vi \right) \\
	&\quad + \tc^2 \Big( c \dc (\SR^2 - \Vi \SLR) + \dc c
	(\SL^2 - \SLR \Vi) \Big),
\end{split}
\end{equation}
which gives
\begin{numcases}{}
	&\!\!\!\!$[\Vi, S_0] + \Vi \SL - \SR \Vi = 0$, \label{odd_S_1}\\
	&\!\!\!\!$\{ S_0, \SR \} + \SR^2 - \Vi \SLR = 0$, \label{odd_S_2}\\
	&\!\!\!\!$\{ S_0, \SL \} + \SL^2 - \SLR \Vi = 0$, \label{odd_S_3}\\
	&\!\!\!\!$[S_0, \SLR] = 0$. \label{odd_S_4}
\end{numcases}

\section{Example of New Objects and Their Relations to Scattering}%
\label{sec:example}

\subsection{SUSY QM in More Detail}%
\label{sub:susy_qm}

Supersymmetric quantum mechanics can be formulated as follows
\begin{equation}\label{susy_qm}
	Q_+ = \D + \D{W}, \quad Q_- = (\D - \D{W})^*, \quad H = \{ Q_+, Q_- \},
\end{equation}
where $\D = \y \dx$ is the de Rham differential, $\y^* = -\dy$ and $\y$ is the
Grassmann variable associated with supersymmetry (for a broad review
see~\cite{cooper1994}).

Let us write down the Hamiltonian and its symmetries explicitly:
\begin{equation}\label{susy_H}
	H = - \dx^2 + W'^2 - W'' (1 - 2 \y \dy),
\end{equation}
\begin{equation}\label{susy_Q}
	Q_+ = \y (\dx + W'), \qquad Q_- = \dy (-\dx + W').
\end{equation}
\begin{remark}
	Both $Q_+$ and $Q_-$ are symmetries of the full Hamiltonian but not of
	any of its subparts.
\end{remark}

Let us choose $S = Q_+$. As we can see, $\SLR \sim \y^2 = 0$, so
equations~\eqref{odd_S_2}-\eqref{odd_S_4} are of no use in this case.
On the other hand, equation~\eqref{odd_S_1} presents an interesting symmetry of
$\Vi$ which, as we have shown earlier, is the K-matrix.

\begin{remark}
	If we were to consider $S = Q_+ + Q_-$, $\SLR$ would surely appear.
\end{remark}

We shall proceed to studying this equation for a particular potential on the
circle of radius $R$ and then acquire the results for the real line by taking
the limit $R \to \infty$.

\subsection{Double $\delta$-potential on $S^1$}%
\label{sub:double_delta_potential_on_s_1_}

The reason for studying the problem on a circle rather than on a line will
become apparent later. For now we say that this case is easier to understand due
to its discrete spectrum. With a continuous spectrum the projection on
$\ker{H_{0,E}}$, represented by two points on the momentum line brings about
unnecessary complications.

As for why we are considering two $\delta$-potentials instead of one, this is
because a single $\delta$-potential can not be produced by a periodic $W'$.

On a circle of radius $R$ we use the coordinate $x \in [-\pi R, \pi R)$. We
consider
\begin{equation}
	W'(x) = \frac{\lambda}{2} \sgn(a^2 - x^2),
\end{equation}
where $\lambda$ and $0 < a < \pi R$ are fixed parameters. According
to~\eqref{susy_H} this produces the potential
\begin{equation}
	\frac{\lambda^2}{4} + (-1)^F \lambda (\delta(x - a) - \delta(x + a)),
\end{equation}
where $F$ is the fermionic number, so $(-1)^F = 1 - 2\y\dy$.

Clearly, as it is now, this potential is inconvenient for a scattering problem
because of the constant shift. We shall send $\frac{\lambda^2}{4}$ to the <<free
part>> of the problem so that the perturbation has a finite support:
\begin{equation}
	H_{0,E} = - \dx^2 - E - \frac{\lambda^2}{4}, \quad
	V(x) = (-1)^F \lambda (\delta(x - a) - \delta(x + a)).
\end{equation}
Similarly, we do the same for the symmetry:
\begin{equation}
	S_0 = \y (\dx - \tfrac{\lambda}{2}), \quad
	S_1 = \y \lambda \theta(a^2 - x^2).
\end{equation}

The spectrum of the free problem is discrete due to periodicity of the wave
functions:
\begin{equation}
	E_n = k_n^2 - \frac{\lambda^2}{4} , \qquad k_n = \frac{n}{R}, \quad n \in
	\mathbb{Z}.
\end{equation}
We fix the ratio $\kappa = \frac{n_0}{R} > 0$ for the energy.
\begin{remark}
	This means that in what follows $R$ will not be an arbitrary positive
	number, but an arbitrary multiple of $\frac{1}{\kappa}$.
\end{remark}
Any level with positive $n$ is double-degenerate, so $\ker{H_{0,E}}$ has
dimension four.
\begin{remark}
	We should not forget to count bosons and fermions separately. As with our
	notation this is done through the Grassmann variable $\y$, our matrices will
	be $2 \times 2$, not $4 \times 4$.
\end{remark}

The normalized wave functions are
\begin{equation}
	\varphi_n (x) = \langle x|n \rangle = \frac{\E^{\I k_n x}}{\sqrt{2 \pi R}}.
\end{equation}
The homological Green's function is simply
\begin{equation}\label{Gh_S1}
	\G = \sum_{|n| \ne n_0} \frac{\ket{n}\bra{n}}{E_{n_0} - E_n}
	 = \sum_{|n| \ne n_0} \frac{\ket{n}\bra{n}}{\kappa^2 - k_n^2}.
\end{equation}
The matrix elements of operators we need are
\begin{align}
	V_{n,m} &= \frac{\I (-1)^F \lambda}{\pi R} \sin{a(k_m-k_n)},
	& G_{\text{h},n,m} &= \frac{\delta_{n,m}}{\kappa^2 - k_n^2}, \\
	S_{0,n,m} &= \y (\I k_n - \tfrac{\lambda}{2}) \delta_{n,m},
	& S_{1,n,m} &= \y \frac{\lambda}{\pi R} \frac{\sin{a(k_m-k_n)}}{k_m-k_n}.
\end{align}
Restriction on-shell is trivial: for some operator $A$ we construct a
$2\times 2$ matrix
\begin{equation}
	A\big|_{\ker{H_{0,E}}} = \pi A i =
	\begin{pmatrix}
		A_{n_0,n_0} & A_{n_0,-n_0} \\
		A_{-n_0,n_0} & A_{-n_0,-n_0}
	\end{pmatrix}.
\end{equation}
\begin{remark}
	Keep in mind that the matrix elements of $A\big|_{\ker{H_{0,E}}}$ may depend
	on $\y$ and $\dy$, as is the case for $V$ and $S$.
\end{remark}
These expressions allow us to calculate $\Vi$, $\SL$ and $\SR$ perturbatively in
$\lambda$. Using a computer algebra system we were able to compute the series up
to $\lambda^3$. The results are given in the appendix.

An important thing to note is that the $n$-th term in the series turns out to be
a polynomial in $R^{-1}$ of degree $n$. Since the supersymmetry holds for
any radius $R$, we have \emph{several} equations for each power of $\lambda$. We
shall denote terms proportional to $\lambda^\alpha R^{-\beta}$ with
$(\alpha,\beta)$. Equation~\eqref{odd_S_1} can now be checked perturbatively for
the first few terms:
\begin{align}
	&(1,1): & &[\Vi^{(1,1)}, S_0^{(0,0)}] = 0, \\
	&(2,1): & &[\Vi^{(2,1)}, S_0^{(0,0)}] + [\Vi^{(1,1)}, S_0^{(1,0)}] = 0, \\
	&(2,2): & &[\Vi^{(2,2)}, S_0^{(0,0)}] + \Vi^{(1,1)} \SL^{(1,1)}
	- \SR^{(1,1)} \Vi^{(1,1)} = 0,\label{l2r2} \\
	&(3,1): & &[\Vi^{(3,1)}, S_0^{(0,0)}] + [\Vi^{(2,1)}, S_0^{(1,0)}] = 0, \\
	&(3,2): & &[\Vi^{(3,2)}, S_0^{(0,0)}] + [\Vi^{(2,2)}, S_0^{(1,0)}]
	\nonumber\\
	& & &+ \Vi^{(1,1)} \SL^{(2,1)} - \SR^{(2,1)} \Vi^{(1,1)} \\
	& & &+ \Vi^{(2,1)} \SL^{(1,1)} - \SR^{(1,1)} \Vi^{(2,1)} = 0, \nonumber\\
	&(3,3): & &[\Vi^{(3,3)}, S_0^{(0,0)}] \nonumber\\
	& & &+ \Vi^{(1,1)} \SL^{(2,2)} - \SR^{(2,2)} \Vi^{(1,1)} \\
	& & &+ \Vi^{(2,2)} \SL^{(1,1)} - \SR^{(1,1)} \Vi^{(2,2)} = 0, \nonumber\\
	&(4,1): & &\ldots\nonumber\\
	&\ldots\nonumber
\end{align}

\subsection{$R \to \infty$ Limit}%
\label{sub:_r_to_infty_limit}

To get more understanding of the equations above it is beneficial to consider
the infinite radius limit. As $R \to \infty$, the discretization becomes
negligible and the problem coincides with that on a line. To recover the
$\delta$-normalization of states we have to multiply each $\ket{n}$ by $\sqrt{2
\pi R}$, which imposes the following relation between matrix elements of
operators:
\begin{equation}\label{S1_to_R}
	A^{\mathbb{R}}(k_n, k_m) = \lim_{R \to \infty} 2 \pi R A^{S^1}_{n,m}.
\end{equation}
Expression~\eqref{Gh_S1} becomes the principal value integral and
\begin{align}
	\Vi^{\mathbb{R}} &= 2\pi R(\Vi^{(1,1)} + \Vi^{(2,1)} + \Vi^{(3,1)} +
	\ldots),\\
	\SL^{\mathbb{R}} &= 2\pi R(\SL^{(1,1)} + \SL^{(2,1)} + \SL^{(3,1)} +
	\ldots),\\
	\SR^{\mathbb{R}} &= 2\pi R(\SR^{(1,1)} + \SR^{(2,1)} + \SR^{(3,1)} +
	\ldots).
\end{align}
The terms given in the appendix agree with the analytical results for
$\Vi^{\mathbb{R}}$, $\SL^{\mathbb{R}}$ and $\SR^{\mathbb{R}}$ on a line, which
we were able to acquire. These formulae are also given in the appendix.

However, when it comes to $S_0$, there is no limit when using
rule~\eqref{S1_to_R}. This is expected, as trying to evaluate
\begin{equation}
	S_0(p,q) = (\I p - \tfrac{\lambda}{2}) \delta(p - q)
\end{equation}
on-shell results in a singularity, and it is well known that $\delta(0) \sim R$
when finite box regularization is used. As a result, there is no simple way to
write, say, equation~\eqref{l2r2} on a line, since the first term is
inderteminate of type <<$0 \times \infty$>>. This result is the main reason
we chose to start with the problem on $S^1$ rather than on $\mathbb{R}$.

\subsection{Concluding Conjecture}%
\label{sub:concluding_conjecture}

We conjecture that it is possible to acquire $(\cdot, \beta)$ terms in $\Qi$ for
$\beta \ge 2$ without referring to the finite radius circle, at least
perturbatively in $\lambda$, as in~\eqref{V_ind}.

\section*{Appendix: Perturbation Series}%

Here we present the terms needed to verify equation~\eqref{odd_S_1} up to
$\lambda^3$ for the double $\delta$-potential on $S^1$.

As in the main text, terms proportional to $\lambda^\alpha R^{-\beta}$ are
denoted with $(\alpha,\beta)$.

\subsection*{$\Vi$ up to $\lambda^3$}%

\begin{equation}
	\Vi^{(1,1)} =
	\begin{pmatrix}
		V_{n_0,n_0} & V_{n_0,-n_0} \\
		V_{-n_0,n_0} & V_{-n_0,-n_0}
	\end{pmatrix} =
	(-1)^F \frac{\lambda \sin{2 a \kappa}}{\pi R} \sigma_2.
\end{equation}

\begin{equation}
	\Vi^{(2,1)} = - \frac{\lambda^2 \sin{2 a \kappa}}{2 \pi R \kappa}
	(\cos{2 a \kappa} + \sigma_1).
\end{equation}

\begin{equation}
	\Vi^{(2,2)} = \frac{\lambda^2 a \sin{2 a \kappa}}{\pi^2 R^2 \kappa}
	\left( \cos{2 a \kappa} - \frac{\sin{2 a \kappa}}{4 a \kappa} + \sigma_1
	\right).
\end{equation}

\begin{equation}
	\Vi^{(3,1)} =
	- (-1)^F \frac{\lambda^3 \sin^3{2 a \kappa}}{4 \pi R \kappa^2} \sigma_2.
\end{equation}

\begin{equation}
	\Vi^{(3,2)} =
	(-1)^F \frac{\lambda^3 (\cos{2 a \kappa} + 4 a \kappa \sin{2 a \kappa})
	\sin^2{2 a \kappa}}{4 \pi^2 R^2 \kappa^3} \sigma_2.
\end{equation}

\begin{equation}
	\Vi^{(3,3)} =
	- (-1)^F \frac{\lambda^3 \big(8 a \kappa \cos{2 a \kappa} + (16 a^2 \kappa^2
	- 1) \sin{2 a \kappa}\big) \sin^2{2 a \kappa}}{16 \pi^3 R^3 \kappa^4}
	\sigma_2.
\end{equation}

\subsection*{$\SL$ up to $\lambda^2$}%

\begin{equation}
	\SL^{(1,1)} = \y \frac{a \lambda}{\pi R} \left( 1 + \frac{\sin{2 a
	\kappa}}{2 a \kappa} \sigma_1 \right).
\end{equation}

\begin{equation}
	\SL^{(2,1)} = \y \frac{\I \lambda^2 a}{2 \pi R \kappa} \Big( (1 -
	\tfrac{\sin{4 a \kappa}}{4 a \kappa}) \sigma_3 + \I (\cos{2 a \kappa} -
	\tfrac{\sin{2 a \kappa}}{2 a \kappa}) \sigma_2 \Big).
\end{equation}

\begin{equation}
\begin{split}
	\SL^{(2,2)} &= \y \frac{\I \lambda^2}{8 \pi^2 R^2 \kappa^3} \Big(
	\cos{4 a \kappa} + 2 a \kappa \sin{4 a \kappa} - 4 a^2 \kappa^2 - 1 \\
	&\qquad\qquad\qquad\qquad~+ \I 4 a \kappa (\sin{2 a \kappa}
	- a \kappa \cos{2 a \kappa}) \sigma_2 \Big).
\end{split}
\end{equation}

\subsection*{$\SR$ up to $\lambda^2$}%

\begin{equation}
	\SR^{(1,1)} = \SL^{(1,1)}, \qquad
	\SR^{(2,1)} = \left( \SL^{(2,1)} \right)^\dagger, \qquad
	\SR^{(2,2)} = \SL^{(2,2)}.
\end{equation}

\subsection*{Analytical Results on $\mathbb{R}$}%

\begin{equation}
	\Vi^{\mathbb{R}} = - \frac{4 \kappa \lambda \sin{2 a \kappa}}{4 \kappa^2 +
	\lambda^2 \sin^2{2 a \kappa}} (\lambda \cos{2 a \kappa} + \lambda \sigma_1 +
	2 \kappa \sigma_2).
\end{equation}

\begin{equation}
\begin{split}
	\SL^{\mathbb{R}} &= \y \frac{4 a \kappa \lambda}{4 \kappa^2 + \lambda^2
	\sin^2{2 a \kappa}} \Bigg( 2\kappa - \I \lambda \left( 1 - \frac{\sin{4 a
	\kappa}}{4 a \kappa} \right) \sigma_3 \\
	&\qquad\qquad+ \frac{\sin{2 a \kappa}}{a} \sigma_1 + \lambda \left( \cos{2 a
	\kappa} - \frac{\sin{2 a \kappa}}{2 a \kappa} \right) \sigma_2 \Bigg).
\end{split}
\end{equation}

\begin{equation}
	\SR^{\mathbb{R}} = \left( \SL^{\mathbb{R}} \right)^\dagger.
\end{equation}

\end{document}